\documentclass{aa}  
\usepackage[T1]{fontenc}
\usepackage{graphicx}
\usepackage{multirow}
\usepackage{amsmath,amssymb,amsfonts}
\usepackage{mathrsfs} 
\usepackage[title]{appendix} 
\usepackage{xcolor}
\usepackage{textcomp} 
\usepackage{manyfoot}
\usepackage{booktabs} 
\usepackage{algorithm} 
\usepackage{algorithmicx}
\usepackage{algpseudocode}
\usepackage{listings}
\usepackage{subcaption}
\usepackage{threeparttable}
\usepackage{tabularx}
\usepackage{txfonts}
\usepackage{hyperref}
\usepackage{placeins}

\newcommand{\singlestoryg}{%
  \text{\fontfamily{cmr}\fontencoding{OT1}\fontshape{it}\selectfont g}%
}

\begin{document}

   \title{Identifying deformed exoplanets with JWST and \textit{Ariel}:\\ A target selection metric for phase curve observations}
   \titlerunning{Identifying deformed exoplanets with JWST and \textit{Ariel}}

   \author{Vanesa Ram\'irez 
    \thanks{Corresponding author: \email{ramirez@strw.leidenuniv.nl}}
    \inst{1}
      \and
      Yamila Miguel
      \inst{1,2}
      \and
      Billy Edwards
      \inst{2}
      \and 
      Babatunde Akinsanmi
      \inst{3,4}
      \and
      Deivy Mercado
      \inst{5}
      }

   \institute{Leiden Observatory, Leiden University,
              PO Box 9513, 2300 RA Leiden, The Netherlands
         \and
             SRON, Space Research Organisation Netherlands, Niels Bohrweg 4, 2333 CA Leiden, The Netherlands
         \and 
            Observatoire astronomique de l’Université de Genève, chemin Pegasi 51, 1290 Versoix, Switzerland
        \and
            School of Physics, University of New South Wales, Kensington, NSW 2052, Australia
        \and 
             Instituto de Física - FCEN, Universidad de Antioquia, Calle 70 No. 52-21, 050001 Medellín, Colombia
             }

   \date{Received 29 July 2026 / Accepted 13 November 2026}
 
  \abstract
   {The shape of an exoplanet encodes fundamental information about its internal structure. For close-in giant planets, strong stellar tidal forces induce measurable deformations that can be probed through high-precision photometry. These deformations are quantified by the fluid Love numbers, which characterise a planet's response to external perturbations and are directly linked to its internal density distribution. In this work, we assessed the potential of JWST and the future \textit{Ariel} mission to constrain the Love numbers of hot Jupiters through full-orbit phase curve observations. To identify the most promising targets for this science case, we introduced a new detection metric, the deformation metric ($\mathcal{D}$), which combines the expected deformation amplitude with the anticipated photometric signal-to-noise ratio. We applied this metric to the \textit{Ariel} target list and identified WASP-33 b, WASP-76 b, and HAT-P-32 b as the three most favourable targets that currently lack Love number measurements. Using WASP-76 b as a benchmark case, we performed end-to-end retrieval simulations and show that a single continuous JWST phase curve observation can constrain the Love number $h_{2f}$ to a precision of $\sigma_{h_{2f}} \approx 0.09$, while a single \textit{Ariel} visit could achieve $\sigma_{h_{2f}} \approx 0.18$. Such precision is likely sufficient for Love number measurements to provide meaningful constraints on planetary interior models and to distinguish between competing interior-structure scenarios, opening a new observational avenue for probing the deep interiors of exoplanets.}

   \keywords{planets and satellites: interiors --
                planets and satellites: fundamental parameters --
                techniques: photometric
               }

   \maketitle
   
\section{Introduction}

Constraining the internal structure of exoplanets remains one of the fundamental challenges of modern planetary science. For most transiting planets, interior inferences rely primarily on bulk mass and radius measurements, occasionally complemented by atmospheric composition constraints \citep{Dorn2015, Dorn2017, Bloot2023, Acuna2024_gastli, vanDijk2025}. However, inferring interior structure from these observables remains an intrinsically degenerate inverse problem. Distinct internal configurations, such as varying core mass fractions, heavy-element enrichment in the envelope, or distinct radial density gradients, can reproduce identical bulk properties \citep[e.g.][]{Valencia2006, Seager2007, Rogers2010, Dorn2015, Nettelmann2021}. Consequently, even for well-characterised hot Jupiters, internal structures often remain poorly constrained. 

Similar interior-structure degeneracies are encountered within the Solar System. Even for Uranus and Neptune, current gravity-field measurements cannot uniquely constrain the distribution of heavy elements in their interiors \citep[e.g.][]{Nettelmann2013, Teanby2020, Morf2025, Ramirez2026}. Such ambiguities are commonly mitigated through measurements of higher-order gravitational moments and tidal Love numbers, which provide complementary constraints on the internal mass distribution. Love numbers inferred from spacecraft tracking and tidal responses have therefore played a central role in constraining the interior structures of Earth, the Moon, Mars, icy satellites, and the giant planets \citep[e.g.][]{Ray2001, Yoder2003, Konopliv2013, Lainey2017, Durante2020}.

Although spatially resolved gravity-field measurements remain beyond reach for exoplanets, Love numbers provide a promising avenue for probing their interiors. Close-in planets are subject to strong tidal forcing from their host stars, leading to deviations from spherical symmetry. In hydrostatic equilibrium, the amplitude of this tidal response is dictated by the degree of central mass concentration and is quantified by the second-order fluid Love numbers, $k_{2f}$ and $h_{2f}$ \citep{Love1911}. These describe, respectively, the perturbation to the gravitational potential and the radial displacement of the surface under an external tidal potential. For a fluid body, these parameters are related by $h_{2f} = 1 + k_{2f}$ \citep{Munk1960}. Their values encode the internal density structure: a strongly centrally condensed object approaches $k_{2f} \rightarrow 0$ ($h_{2f}\rightarrow 1$), whereas a fully homogeneous body yields $k_{2f}=1.5$ ($h_{2f}=2.5$) \citep{Kramm2011, Padovan2018}. Because $k_{2f}$ is highly sensitive to heavy-element gradients and core masses \citep[e.g.][]{Kramm2011, Buhler2016, vanDijk2025}, measuring this shape distortion provides a powerful complementary constraint that can help break long-standing degeneracies in giant planet interiors.

Despite its diagnostic power, measuring Love numbers in exoplanets is observationally challenging. Two principal approaches have been proposed. The first exploits orbital dynamics, including apsidal precession and secular interactions in multi-planet systems, where the orbital evolution depends directly on the planetary $k_{2f}$ \citep[e.g.][]{Buhler2016, Csizmadia2019, Bernabo2024, Nediyedath2026}. The second relies on detecting photometric signatures of tidal deformation, to measure $h_{2f}$, in transit light curves and phase curves \citep{Correia2014a, Akinsanmi2019, Hellard2019, Akinsanmi2024_theory, Akinsanmi2024_wasp12}. In this photometric approach, the tidally distorted planet's projected area varies with orbital phase, producing subtle but characteristic deviations from spherical models. These photometric signals are small, typically at the level of tens to a few hundred parts per million (ppm), and can be partially degenerate with other phase curve components. As a result, only a small number of exoplanets currently possess published Love-number measurements or meaningful constraints, summarised in Table~\ref{table:love_numbers}. Although these pioneering studies demonstrate observational feasibility, uncertainties remain large in most cases, and measurements are restricted to a few exceptional systems.

\begin{table*}[!t]
\centering
\caption{Summary of currently published exoplanet Love number measurements.}
\label{table:love_numbers}

\begin{threeparttable}
\begin{tabularx}{\textwidth}{l c c X l}
\toprule
\toprule
Planet & Type & Measurement & Method & Reference \\
\midrule

\multirow{2}{*}{WASP-12 b}
& $h_{2f}$ & $1.55^{+0.45}_{-0.49}$
& Phase curve shape analysis
& \citet{Akinsanmi2024_wasp12} \\

& $k_{2f}$ & $0.63 \pm 0.089$
& Apsidal precession / transit timing
& \citet{Biswas2026} \\

\addlinespace

WASP-103 b
& $h_{2f}$ & $1.59^{+0.45}_{-0.53}$
& Transit shape analysis
& \citet{Barros2022} \\

\addlinespace

\multirow{2}{*}{WASP-18 b}
& $k_{2f}$ & $0.62^{+0.55}_{-0.19}$
& Apsidal precession (radial velocity + timing)
& \citet{Csizmadia2019} \\

& $k_{2f}$ & $0.62199 \pm 0.0011$
& Apsidal precession (radial velocity + timing)
& \citet{Nediyedath2026} \\

\addlinespace

\multirow{2}{*}{HAT-P-13 b}
& $k_{2f}$ & $0.31^{+0.08}_{-0.05}$
& Dynamical modelling
& \citet{Buhler2016} \\

& $k_{2f}$ & $0.81 \pm 0.10$
& Dynamical modelling
& \citet{Hardy2017} \\

\addlinespace

\multirow{2}{*}{WASP-19 b}
& $k_{2f}$ & $0.20^{+0.02}_{-0.03}$
& Apsidal precession / timing
& \citet{Bernabo2024} \\

& $k_{2f}$ & $0.107 \pm 0.08$
& Apsidal precession / timing
& \citet{Rajkumar2026} \\

\addlinespace

WASP-121 b
& $h_{2f}$ & $1.39^{+0.71}_{-0.81}$
& Transit shape analysis (tentative)
& \citet{Hellard2020} \\

\bottomrule
\end{tabularx}
\end{threeparttable}
\begin{tablenotes}[flushleft]
\item \textbf{Notes.} The reported values include both the radial fluid Love number ($h_{2f}$), typically derived from direct shape deformation modelling, and the potential fluid Love number ($k_{2f}$), constrained via orbital dynamics and apsidal precession. Values derived from apsidal precession or orbital dynamics technically measure the potential Love number ($k_2$). Under the standard assumption that these hot Jupiters are fluid bodies in hydrostatic equilibrium, we report these measurements as their equivalent fluid Love number ($k_{2f}$, where $h_{2f} = 1 + k_{2f}$). The scarcity and large uncertainties of current measurements underscore the strong motivation to identify new optimal targets and expand the sample of observationally constrained planetary Love numbers.
\end{tablenotes}
\end{table*}

The advent of ultra-high-precision space-based photometry now opens a new regime for deformation studies. The \textit{James Webb} Space Telescope (JWST) delivers unprecedented signal-to-noise ratios, enabling detailed characterisation of hot-Jupiter atmospheres and their geometric signatures in single-visit full-orbit phase curves. Rather than relying on isolated proof-of-concept detections, JWST offers an opportunity to conduct systematic searches for tidally deformed planets. In the next decade, the forthcoming \textit{Ariel} mission will provide a natural complement, expanding this parameter space to a significantly broader catalogue using its dedicated spectrophotometric monitoring capabilities.

In this work, we bridge the gap between theoretical planetary deformation and practical observation strategies to assess the capability of JWST and \textit{Ariel} to reliably measure exoplanetary Love numbers. We first revisit the formalism of tidal deformation and its photometric signatures (Sect.~\ref{sec:theory}). We then introduce the deformation metric, a novel targeting criterion combining the theoretical amplitude of physical deformation with the expected photometric signal-to-noise ratio of the system to systematically prioritise the most promising candidates (Sect.~\ref{sec:metric}). Adopting WASP-76 b as a benchmark, we performed rigorous end-to-end forward modelling and Bayesian retrievals for JWST and \textit{Ariel} to determine the achievable observational precision on the fluid Love number (Sect.~\ref{sec:simulations_wasp76b}). We then discuss the observational strategies required to achieve such constraints (Sect.~\ref{sec:discussion}). We show that single, continuous full-orbit phase curves provide a more efficient and robust route to measuring planetary deformation than either transit-only observations or extensive multi-visit campaigns, for both JWST and \textit{Ariel}. Finally, in Sect.~\ref{sec:conclusions}, we summarise our main findings and their implications for the future of exoplanet characterisation.

\section{Theory and formalism of deformation signatures}
\label{sec:theory}

\subsection{The tidal potential and Love numbers}

Close-in giant planets, particularly ultra-hot Jupiters, reside in extreme gravitational environments in which tidal interactions with their host stars drive them away from spherical symmetry. To first order, the radial displacement of the planetary surface, $\Delta R(\theta, \varphi)$, relative to its unperturbed spherical radius $R_0$, is proportional to the external perturbing potential $V_{\rm ext}(\theta, \varphi)$. In linear tidal theory, this response is quantified by the fluid radial Love number, $h_{2f}$ \citep{Love1911, Correia2014a, Akinsanmi2019}:

\begin{equation}
\Delta R(\theta, \varphi) = h_{2f} \frac{V_{\rm ext}(\theta, \varphi)}{g_0},
\end{equation}
\noindent
where $g_0$ is the surface gravity. For a fluid body in hydrostatic equilibrium, this radial deformation is fundamentally linked to the potential Love number through $h_{2f} = 1 + k_{2f}$ \citep{Munk1960}.

While $k_{2f}$ characterises the perturbation to the planet's gravitational field, $h_{2f}$ directly dictates the observable geometric distortion of the planetary surface. Because this response is governed by the internal mass distribution \citep{Kramm2011, Padovan2018}, an empirical measurement of $h_{2f}$ provides a powerful observational constraint that helps mitigate degeneracies in planetary interior models.

\subsection{Planet shape model}

Tidally deformed planets can be approximated as triaxial ellipsoids, elongated along the substellar-antistellar axis and compressed along the polar axis \citep{Correia2014a}. We define the principal semi-axes, $R_x$, $R_y$, and $R_z$, to be aligned with the star-planet axis, the orthogonal axis within the orbital plane, and the polar rotation axis, respectively. The volumetric radius of the planet is then given by $R_{\rm v} = (R_x R_y R_z)^{1/3}$. Following the analytical formalism developed by \citet{Correia2014a} and \citet{Akinsanmi2019}, the principal axes are related, to first order, by

\begin{align}
R_y &= R_{\rm v}\left(1-\frac{2}{3}\mathcal{A}_{\rm def}\right), \\
R_x &= R_y\left(1+3\mathcal{A}_{\rm def}\right), \\
R_z &= R_y\left(1-\mathcal{A}_{\rm def}\right),
\end{align}
\noindent
where $\mathcal{A}_{\rm def}$ is a dimensionless parameter that quantifies the tidal deformation. Under the assumption of a fluid body in hydrostatic equilibrium, $\mathcal{A}_{\rm def}$ is directly proportional to the radial Love number, $h_{2f}$:

\begin{equation}
\mathcal{A}_{\rm def} = \frac{h_{2f}}{2} \frac{M_{\rm s}}{M_{\rm p}} \left(\frac{R_{\rm v}}{a}\right)^3,
\label{eq:Adef}
\end{equation}
\noindent
where $M_{\rm s}$ and $M_{\rm p}$ are the stellar and planetary masses, and $a$ is the semi-major axis. This scaling relation shows that planets with large radii orbiting massive host stars at small orbital separations are optimal targets for measuring shape deformation due to the extreme tidal forces they experience. The detectability additionally depends on the achievable photometric precision and phase curve coverage.

As the synchronously rotating, triaxial planet orbits its host star, the observer sees a continuously changing projected cross-sectional area. Defining the orbital phase $\phi$ such that mid-transit occurs at $\phi=0$, the projected area for a system with orbital inclination $i$ is given by \citep{Leconte2011, Akinsanmi2024_theory}

\begin{equation}
A(\phi) = \pi \sqrt{R_x^2 R_y^2 \cos^2 i + R_z^2 \sin^2 i \left( R_x^2 \sin^2\phi + R_y^2 \cos^2\phi \right) }.
\label{eq:projected_area}
\end{equation}
\noindent
To first order in $\mathcal{A}_{\rm def}$, Eq.~(\ref{eq:projected_area}) can be written as

\begin{equation}
A(\phi)
\simeq
\pi R_{\rm v}^2
\left[
1
+
\mathcal{A}_{\rm def}
\left(
-\frac56
+\frac52\cos^2 i
-\frac32\sin^2 i \cos2\phi
\right)
\right].
\label{eq:projected_area_linear}
\end{equation}
\noindent
The first-order expansion shows explicitly that the projected area contains a symmetric $\cos2\phi$ harmonic with semi-amplitude (3/2)$\mathcal{A}_{\rm def} \sin^2 i$. This modulation constitutes the observable photometric signature of tidal deformation and forms the basis of the target-selection metric introduced in Sect.~\ref{sec:metric}.

\subsection{Photometric signatures in light curves}
\label{sec:photometric_signature}

Observational efforts to constrain planetary shape have historically focused on primary transits \citep[e.g.][]{Barnes2003, Carter2010, Zhu2014, Biersteker2017, Barros2022}. As a deformed planet crosses the stellar disc, its non-circular projected limb marginally alters the transit depth relative to an equivalent sphere and introduces subtle deviations during ingress and egress \citep{Leconte2011, Correia2014a, Akinsanmi2019, Hellard2019}. However, these transit-derived signatures are exceptionally subtle and remain strongly degenerate with stellar limb-darkening coefficients, as well as with planetary and orbital parameters such as the planetary radius, impact parameter, and semi-major axis \citep[e.g.][]{Hellard2019, Akinsanmi2019, Barros2022}.

To overcome the limitations and degeneracies inherent to transit-only measurements, recent studies have advocated the use of full-orbit phase curves to characterise tidal deformation \citep{Akinsanmi2024_theory, Akinsanmi2024_wasp12}. Following their formalism, the total normalised flux of the system, $F(\phi)$, as a function of the orbital phase $\phi$, can be modelled as the sum of the stellar and planetary components, modulated by the transit and occultation functions:

\begin{equation}
F(\phi) = F_{\rm tra}(\phi) F_{\rm s}(\phi) + F_{\rm occ}(\phi) F_{\rm p}(\phi),
\end{equation}
\noindent
where $F_{\rm tra}$ and $F_{\rm occ}$ describe the primary transit and secondary eclipse (occultation), respectively.

The out-of-transit stellar flux, $F_{\rm s}(\phi)$, is normalised to unity but exhibits phase-dependent variations due to the gravitational interaction with the planet. These arise from stellar ellipsoidal variation ($F_{\rm EV}$), induced by the planet's tidal pull, and Doppler beaming ($F_{\rm DB}$), arising from the radial motion of the host star:

\begin{equation}
F_{\rm s}(\phi) = 1 + F_{\rm EV}(\phi) + F_{\rm DB}(\phi),
\end{equation}
\noindent
where the individual components are modelled as

\begin{align}
F_{\rm EV}(\phi) &= A_{\rm EV}(1 - \cos 2\phi), \\
F_{\rm DB}(\phi) &= A_{\rm DB} \sin \phi.
\end{align}
\noindent
Here, $A_{\rm EV}$ and $A_{\rm DB}$ are the semi-amplitudes of the stellar ellipsoidal and Doppler beaming effects, respectively. The semi-amplitude of the stellar ellipsoidal variation is given by \citep{Esteves2013}

\begin{equation}
A_{\rm EV} = \alpha_{\rm EV} \frac{M_{\rm p}}{M_{\rm s}} \left(\frac{R_{\rm s}}{a}\right)^3 \sin^2 i,
\label{eq:A_EV}
\end{equation}
\noindent
with $\alpha_{\rm EV}$ the passband-dependent ellipsoidal coefficient:

\begin{equation}
\alpha_{\rm EV} = 0.15 \frac{(15+u)(1+g)}{3-u},
\label{eq:alpha_EV}
\end{equation}
\noindent
and $u$ and $g$ the linear limb-darkening and gravity-darkening coefficients, respectively. Similarly, the star's reflex motion induces a Doppler beaming effect, which modifies the observed flux due to relativistic beaming and the Doppler shift of the stellar spectrum. The semi-amplitude of the Doppler beaming effect is given by \citep{Shporer2017}

\begin{equation}
A_{\rm DB} = \alpha_{\rm DB}4 \frac{K}{c},
\label{eq:A_DB}
\end{equation}
\noindent
where $K$ is the radial velocity semi-amplitude, $c$ is the speed of light, and $\alpha_{\rm DB}$ is the beaming coefficient evaluated at the reference frequency $\nu$ of the observing passband:

\begin{equation}
\alpha_{\rm DB} = \frac{1}{4}\frac{x e^x}{(e^x - 1)}, \quad \text{with} \quad x = \frac{h \nu}{k_{\rm B} T_{\rm eff}},
\end{equation}
\noindent
in which $h$ is the Planck constant, $k_{\rm B}$ is the Boltzmann constant, and $T_{\rm eff}$ is the stellar effective temperature.

For a standard, spherical planet, the planetary flux contribution $F_{\rm p}(\phi)$ represents the atmospheric phase variation, which is a combination of thermal emission and reflected light. This is commonly modelled as a sinusoidal function parametrised by the maximum ($F_{\rm max}$) and minimum ($F_{\rm min}$) planetary fluxes, alongside an atmospheric hotspot offset $\delta$ caused by advective winds:

\begin{equation}
F_{\rm p}(\phi) = (F_{\rm max} - F_{\rm min}) \frac{1 - \cos (\phi + \delta)}{2} + F_{\rm min}.
\end{equation}
\noindent
The corresponding dayside and nightside fluxes, $F_{\rm d}$ and $F_{\rm n}$, are defined as the values of $F_{\rm p}(\phi)$ at $\phi=\pi$ and $\phi=0$, respectively.

For a tidally deformed planet, however, the planetary flux is additionally modulated by the varying projected cross-sectional area. The emitted and reflected flux of the deformed planet, $F_{\rm p, def}(\phi)$, is scaled by the ratio of the instantaneous projected area $A(\phi)$ (defined in Eqs.~\ref{eq:projected_area} and \ref{eq:projected_area_linear}) to the projected area at conjunction $A(0)$:

\begin{equation}
F_{\rm p, def}(\phi) = F_{\rm p}(\phi) \frac{A(\phi)}{A(0)}.
\label{eq:flux_deformed}
\end{equation}
\noindent
This geometric scaling introduces a symmetric $\cos(2\phi)$ harmonic component into the planetary phase curve. The characteristic symmetry of this deformation-induced signal is determined by geometric projection effects, while its amplitude depends on the degree of tidal deformation and the underlying planetary flux. In contrast, atmospheric thermal emission ($F_{\rm p}(\phi)$) exhibits a strong wavelength dependence, and its peak is typically shifted away from the secondary eclipse ($\delta \neq 0$). Consequently, multi-band photometry provides a means of disentangling the symmetric deformation signal from asymmetric, wavelength-dependent atmospheric variability. For a detailed derivation of these flux components and their implementation in phase curve modelling, we refer the reader to the detailed formalisms presented in \citet{Akinsanmi2024_theory} and \citet{Akinsanmi2024_wasp12}.

\section{The deformation metric ($\mathcal{D}$)}
\label{sec:metric}

\subsection{Definition}

Given the limited observing time available on flagship space observatories, efficient target selection is critical. To systematically identify the most promising targets for detecting tidal deformation through phase curve observations, we introduce the deformation metric ($\mathcal{D}$). This metric is designed to rank targets according to the expected detectability of the deformation-induced phase curve signal. 

The observable signature of tidal deformation is the phase-dependent modulation of the planetary flux (Eq.~\ref{eq:flux_deformed}), which arises from the varying projected area of the planet (Eqs.~\ref{eq:projected_area} and \ref{eq:projected_area_linear}). The corresponding semi-amplitude scales as $\mathcal{A}_{\rm def} \sin^2 i$, indicating that the deformation signal depends not only on the intrinsic degree of tidal distortion, but also on the orbital inclination. The deformation metric therefore combines this observable deformation amplitude with a proxy for the photometric signal-to-noise ratio (S/N) achievable in a full-orbit phase curve observation. We define

\begin{equation}
\mathcal D
=
\mathcal N
\,
\mathcal A_{\rm def}
\sin^2 i
\,
{\rm S/N},
\end{equation}
\noindent
where $\mathcal{N} = 1.43 \times 10^{6}$ is a normalisation constant chosen such that the highest-ranked target in our sample has a metric value of $\mathcal{D} = 1$. 

\begin{figure}[!t]
\centering 
\includegraphics[width=0.48\textwidth]{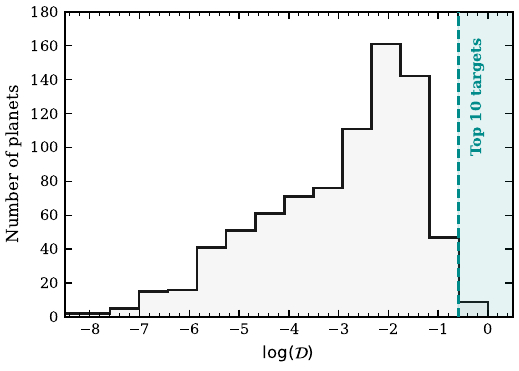}
\caption{Distribution of the deformation metric ($\mathcal{D}$) across the \textit{Ariel} target list. The histogram is plotted on a logarithmic scale to accommodate the wide range of metric values. The vertical dashed cyan line indicates the lower threshold for the ten highest-ranked candidates (Table~\ref{table:candidates}). The long tail toward lower values illustrates that while the vast majority of the population exhibits negligible observable distortion, a distinct, small subset of targets possesses exceptionally strong signals suitable for phase curve characterisation.}
\label{fig:D_histogram}
\label{fig:D_distribution}
\end{figure}

To estimate the observational S/N relevant for phase curve measurements, we followed the formalism of \citet{Charnay2022}. We assumed that the planetary phase curve is sinusoidal, with no phase offset, and that the observation is divided into one-hour bins. Under these assumptions, the detectability of the phase curve signal scales with the thermal planet-to-star flux contrast and with the square root of the total observing time accumulated over the orbit. The absolute S/N depends on the instrument-specific noise for a one-hour integration ($\sigma_{\mathrm{1h}}$). However, the purpose of $\mathcal{D}$ is to provide a relative ranking of targets rather than an instrument-dependent prediction of detectability. We therefore restricted our analysis to the photon-noise-dominated regime and assumed a uniform observing configuration across the sample. In this limit, telescope-dependent quantities such as collecting area and throughput enter only as global proportionality constants and do not affect the relative ranking. Following the emission spectroscopy metric \citep[ESM;][]{Kempton2018}, we accounted for target-to-target variations in photon noise through a brightness scaling proportional to $10^{-m_{\rm K}/5}$, where $m_{\rm K}$ is the apparent $K$-band magnitude. The relative S/N defining our metric is therefore

\begin{equation}
\mathrm{S/N} \approx \frac{1}{2} 10^{-m_{\rm K}/5} \frac{B_{4.5}(T_{\rm p})}{B_{4.5}(T_{\rm eff})}\left(\frac{R_{\rm p}}{R_{\rm s}}\right)^2 \left( \frac{P}{2} \right)^{\frac{1}{2}}.
\end{equation}
\noindent
Here, $B_{4.5}$ denotes the Planck function evaluated at a reference wavelength of $4.5\,\mu$m, representative of the infrared wavelength range relevant to the JWST and \textit{Ariel} configurations considered in this study. $T_{\rm p}$ is the planetary equilibrium temperature computed assuming no heat redistribution, with a Bond albedo of $A_{\rm B}=0.1$ for planets with $T_{\rm p}>700$ K and $A_{\rm B}=0.3$ for planets with $T_{\rm p}<700$ K, following \citet{Charnay2022}. $T_{\rm eff}$ is the stellar effective temperature, $R_{\rm p}$ and $R_{\rm s}$ are the planetary and stellar radii, and $P$ is the orbital period in hours.

\subsection{Candidate selection}
\label{sec:candidate_selection}

\begin{figure}[!t]
\centering 
\includegraphics[width=0.5\textwidth]{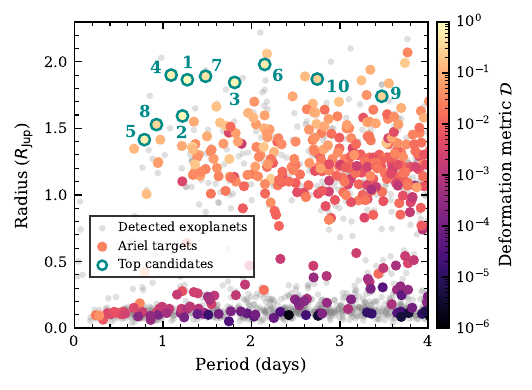}
\caption{Distribution of the analysed exoplanet population in the planetary radius versus orbital period plane. Background grey circles represent the broader population of known transiting exoplanets from the NASA Exoplanet Archive. The \textit{Ariel} target list is superimposed and colours-coded by the value of the deformation metric ($\mathcal{D}$). The ten highest-ranked targets for phase curve deformation studies are highlighted with cyan open circles and numbered from 1 to 10 in order of descending $\mathcal{D}$ value (Table~\ref{table:candidates} for their detailed physical and orbital parameters). The optimal candidates clearly cluster in the upper-left quadrant, demonstrating the metric's strong selection towards highly inflated giant planets on ultra-short-period orbits.}
\label{fig:candidates}
\end{figure}

We calculated the deformation metric for a comprehensive sample of known transiting exoplanets. In this work, we focused specifically on the potential targets for the \textit{Ariel} mission (\textit{Ariel} mission candidate sample\footnote{\url{https://github.com/arielmission-space/Mission_Candidate_Sample}, v. 2025-08-18.}; \citealt{Edwards2019, mugnai_arielrad, Edwards2022}), which is compiled from currently known exoplanets and TESS planet candidates. The release used in this study comprised 814 targets. These systems possess well-characterised stellar and planetary parameters tailored for future space-based atmospheric surveys, making them an ideal statistical sample for our population study. All parameters required to evaluate the metric were sourced directly from this catalogue, using the reported 2MASS $K_{\rm s}$ magnitudes as a proxy for the $K$-band apparent magnitude, $m_{\rm K}$. By definition, $\mathcal{D}$ naturally favours planets with ultra-short orbital periods (maximising the tidal potential) and bright host stars (minimising photometric noise).

To evaluate $\mathcal{D}$ across the \textit{Ariel} target list, we used the planetary radii reported in the mission catalogue. These literature values are primarily derived from standard transit models that assume a spherical geometry and therefore probe the planet's cross-section perpendicular to the star-planet axis. For our population-level calculations, we adopted the reported radius, $R_{\rm p}$, as a first-order proxy for the volumetric radius, $R_{\rm v}$. We note that for the most severely distorted ultra-short-period planets, this approximation systematically underestimates the true volumetric radius ($R_{\rm p} < R_{\rm v}$) and, consequently, yields a conservative lower limit on $\mathcal{D}$. Accurately deriving $R_{\rm v}$ for hundreds of systems would require iterative fits with full triaxial shape models while simultaneously resolving complex parameter degeneracies, such as those involving stellar limb-darkening coefficients. Because this level of detailed, system-specific modelling is beyond the scope of a population-level ranking analysis, we evaluated the metric directly using the planetary radii reported in the catalogue.

Furthermore, because observational constraints on planetary Love numbers remain scarce, we computed the deformation amplitude assuming a fiducial fluid Love number of $h_{2f}~=~1.5$ for all candidates. This value is representative of fluid giant planets and is comparable to Jupiter's measured value of $h_{2f}~\approx~1.565$ \citep{Durante2020}, providing a uniform basis for comparing targets across the sample. Although the intrinsic Love number is expected to vary from planet to planet according to its internal structure and composition, adopting a common fiducial value avoids introducing highly unconstrained, system-specific assumptions and provides a consistent framework for the population-level target ranking considered in this work.

The distribution of $\mathcal{D}$ (Fig.~\ref{fig:D_distribution}) peaks near $\mathcal{D} \sim 10^{-2}$ and exhibits a long tail extending towards lower values. This suggests that observable shape deformation is likely confined to only a small subset of the planetary population, whereas the vast majority of systems do not provide favourable conditions for such measurements. We focused our analysis on the ten highest-ranked systems according to $\mathcal{D}$. Their relevant orbital and physical properties are summarised in Table~\ref{table:candidates}.

Figure~\ref{fig:candidates} maps the relevant parameter space of planetary radii versus orbital periods. The background grey points represent the full population of known transiting exoplanets\footnote{NASA Exoplanet Archive: \url{https://exoplanetarchive.ipac.caltech.edu}}. The \textit{Ariel} target list is superimposed and colour-coded according to the values of $\mathcal{D}$. The ten most promising systems for phase curve deformation studies, ranked by $\mathcal{D}$, are highlighted with cyan circles. The highest-ranked targets cluster among inflated giant planets on ultra-short-period orbits, indicating that planetary radius and proximity to the host star are the primary drivers of detectability. The theoretical deformation amplitude scales as $(R_{\rm p}/a)^3$, or equivalently $(R_{\rm p}^3/P^2)$, implying that larger planets and more compact orbits experience substantially stronger shape distortions. However, the photometric signal-to-noise ratio scales approximately as $(R_{\rm p}^2 P^{1/2})$, partially compensating for the weaker deformation signal in longer-period systems. Consequently, although short-period planets are generally favoured, several systems with orbital periods of a few days remain highly ranked because of their inflated radii. The optimal targets therefore represent a balance between these competing scaling relations. Departures from these broad trends are driven by secondary dependences, including the planet-to-star mass ratio, stellar radius, host-star brightness in the $K$ band, and the thermal flux contrast, which introduce the scatter visible in Fig.~\ref{fig:candidates}. The dependence of $\mathcal{D}$ on the main planetary, orbital, and stellar properties of the \textit{Ariel} target population is explored in more detail in Appendix~\ref{app:deformation_metric_dependence}.

\newcolumntype{Y}{>{\centering\arraybackslash}X}
\begin{table*}[!t]
\caption{Ten highest-ranked candidates for phase curve deformation studies, ordered by their descending deformation metric ($\mathcal{D}$).}
\label{table:candidates}
\centering
\begin{threeparttable}
\renewcommand{\arraystretch}{1.3}
\begin{tabularx}{1.0\textwidth}{YlYYYYYYY}
\toprule
\toprule
Rank & Planet & $P$ (d) & $R$ ($R_{\rm Jup}$) & $M$ ($M_{\rm Jup}$) & $T_{\rm eq}$ (K) & $i$ (deg) & $K_s$ (mag) & $\mathcal{D}$ \\
\midrule
1  & WASP-121 b$^{\rm a}$ & 1.27 & $1.87^{+0.04}_{-0.04}$ & $1.18^{+0.06}_{-0.06}$ & 2297 & 87.60 & 9.37  & 1.00 \\
2  & WASP-33 b  & 1.22 & $1.59^{+0.07}_{-0.07}$ & $2.09^{+0.14}_{-0.14}$ & 2712 & 86.63 & 7.47  & 0.93 \\
3  & WASP-76 b  & 1.81 & $1.85^{+0.05}_{-0.05}$ & $0.91^{+0.04}_{-0.04}$ & 2172 & 88.50 & 8.24  & 0.78 \\
4  & WASP-12 b$^{\rm a}$ & 1.09 & $1.90^{+0.06}_{-0.06}$ & $1.47^{+0.08}_{-0.07}$ & 2514 & 83.37 & 10.19 & 0.70 \\
5  & WASP-19 b$^{\rm a}$ & 0.79 & $1.42^{+0.04}_{-0.05}$ & $1.15^{+0.08}_{-0.08}$ & 2058 & 79.08 & 10.48 & 0.63 \\
6  & HAT-P-32 b & 2.15 & $1.98^{+0.05}_{-0.05}$ & $0.68^{+0.11}_{-0.10}$ & 1788 & 88.98 & 9.99  & 0.61 \\
7  & KELT-9 b & 1.48 & $1.89^{+0.06}_{-0.05}$ & $2.88^{+0.84}_{-0.84}$ & 3945 & 86.79 & 7.48  & 0.45 \\
8  & WASP-103 b$^{\rm a}$ & 0.93 & $1.53^{+0.07}_{-0.05}$ & $1.49^{+0.09}_{-0.09}$ & 2441 & 86.30 & 10.77 & 0.31 \\
9  & KELT-20 b & 3.47 & $1.74^{+0.07}_{-0.07}$ & $0.97^{+0.81}_{-0.44}$ & 2201 & 86.12 & 7.42  & 0.26 \\
10 & HAT-P-70 b & 2.74 & $1.87^{+0.15}_{-0.10}$ & $0.92^{+0.70}_{-0.40}$ & 2485 & 96.50 & 8.96  & 0.25 \\
\bottomrule
\end{tabularx}
\begin{tablenotes}
\item \textbf{Notes.} Orbital period, planetary radius and mass, inclination, and host-star $K_{\rm s}$-band magnitude were adopted from the \textit{Ariel} mission candidate sample catalogue. The planetary equilibrium temperature was computed assuming no heat redistribution and adopting a Bond albedo of $A_{\rm B}=0.1$ for planets hotter than 700\,K and $A_{\rm B}=0.3$ otherwise.

$^{\rm a}$ Planet with a previously published Love number measurement
(Table~\ref{table:love_numbers}).
\end{tablenotes}
\end{threeparttable}
\end{table*}

Ultra-hot Jupiters with inflated radii and extreme levels of stellar irradiation, such as WASP-121\,b, WASP-33\,b, WASP-76\,b, WASP-12\,b, WASP-19\,b, HAT-P-32\,b, KELT-9\,b, WASP-103\,b, KELT-20\,b, and HAT-P-70\,b, exhibit the largest deformation signals and consequently attain the highest metric values. Within this top tier, WASP-121 b, WASP-12 b, WASP-19 b, and WASP-103 b have already been the subject of pioneering Love number measurements (Table~\ref{table:love_numbers}). However, the measurement for WASP-121 b remains tentative, while the reported constraints for WASP-12 b and WASP-103 b remain highly uncertain, with relative uncertainties of approximately 30\% and 80\% in $h_{2f}$ and $k_{2f}$, respectively. These uncertainties are currently too large to place meaningful constraints on their interior structures \citep{vanDijk2025}. Future high-precision phase curve observations of these systems will therefore be essential for refining these pioneering measurements.

Among this sample, WASP-76\,b emerges as a particularly compelling target for an in-depth case study. It combines a highly inflated radius ($R_{\rm p} \sim 1.85\,R_{\rm Jup}$) and a near-Jovian mass ($M_{\rm p} \sim 0.91\,M_{\rm Jup}$), resulting in a low bulk density, an ultra-short orbital period of approximately 1.8 days, and a bright host star with a $K_{\rm s}$-band magnitude of $\sim 8.24$. Crucially, its full-orbit phase curve has been extensively observed with CHEOPS, TESS, and \textit{Spitzer} \citep[e.g.][]{May_wasp76b, Fu2021, Demangeon2024}, and has recently been targeted by JWST/NIRSpec. Although we did not attempt to directly retrieve the fluid Love number from these archival observations, we used atmospheric parameters derived from previous analyses to construct the theoretical baseline for our end-to-end forward models. Anchoring our shape models to these observationally inferred atmospheric properties ensures that our simulated observations remain physically realistic. We therefore adopted WASP-76 b as our benchmark system for evaluating the capability of JWST and the future \textit{Ariel} mission to detect the phase curve signature of tidal deformation and constrain Love numbers in tidally distorted exoplanets.

\section{Case study: WASP-76\,b}
\label{sec:simulations_wasp76b}

To assess the precision with which the radial Love number $h_{2f}$ can be constrained using next-generation space observatories, we conducted a detailed forward-modelling and retrieval experiment using WASP-76 b as a benchmark target. As no empirical measurement of tidal deformation is currently available for WASP-76 b, we injected a fiducial deformation signal corresponding to $h_{2f}=1.5$, comparable to Jupiter's value.  The adopted stellar and planetary parameters, together with the derived photometric quantities used to model the system, are listed in Table~\ref{table:wasp76b_params}. 

\subsection{Forward model}
\label{sec:forward_model}

We generated synthetic full-orbit phase curves incorporating contributions from the primary transit, secondary eclipse, atmospheric thermal emission, reflected light, planetary ellipsoidal modulation, and stellar phase variations (ellipsoidal variation and Doppler beaming), as detailed in Sect.~\ref{sec:photometric_signature}. The light curves were computed using the publicly available tool \texttt{ellc} \citep{Maxted2016}, coupled with the \texttt{tidef} module \citep{Akinsanmi2024_theory, Akinsanmi2024_wasp12}\footnote{\url{https://github.com/tundeakins/tidef}}. The tidal deformation signal was calculated self-consistently such that the phase-dependent projected area of the triaxial ellipsoid naturally  modulates the planetary flux.

To assess the importance of continuous full-orbit coverage, we performed retrieval analyses using both complete phase curve datasets and transit-only subsets for JWST and \textit{Ariel}. Random Gaussian noise was injected into the synthetic light curves using the instrument-specific noise estimates described in Sect.~\ref{sec:instrument_config} and summarised in Table~\ref{table:instruments}. To reduce sensitivity to individual noise realisations, we repeated the analysis for ten independent noise realisations and combined the resulting posterior distributions. 

To evaluate the impact of repeated observations with \textit{Ariel}, we additionally generated datasets corresponding to 1, 3, 5, and 10 co-added visits. For these stacked datasets, we assumed that the observations are dominated by white noise, such that the photometric uncertainty decreases as $1/\sqrt{N}$, where $N$ is the number of visits. In contrast, for JWST we considered only a single visit, reflecting both its superior single-epoch photometric precision and the highly competitive nature of its observing time allocation.

\subsection{Calculation of stellar and photometric parameters}
\label{sec:parameter_calculation}

To generate accurate forward models for the different instrumental passbands, we calculated the wavelength-dependent stellar parameters following the formalism described in Sect.~\ref{sec:photometric_signature}. We adopted the fundamental stellar and planetary parameters of the WASP-76 system from \citet{Fu2021}, listed in Table~\ref{table:wasp76b_params}. The stellar limb-darkening coefficients required to evaluate $\alpha_{\rm EV}$ were computed for each JWST and \textit{Ariel} passband using the \texttt{LDTk} package \citep{Husser2013}, which generates stellar intensity profiles based on the PHOENIX atmosphere models. We derived both the linear ($u$) and power-2 ($c, \alpha$) limb-darkening coefficients using a Monte Carlo approach with 10\,000 samples to robustly propagate the uncertainties in the stellar parameter. The corresponding gravity-darkening coefficients ($g$) were subsequently calculated using the \texttt{ellc} package, which interpolates the tabulated coefficients from \citet{Claret2011}. Given the infrared wavelength coverage of our simulated observations, these coefficients were derived using the \textit{Spitzer}/IRAC $s4.5\,\mu\text{m}$ (Channel~2) bandpass as a representative reference.

\begin{table}[!t]
\centering
\caption{Stellar and planetary parameters adopted for the WASP-76~b phase curve simulations.}
\label{table:wasp76b_params}
\renewcommand{\arraystretch}{1.3}
\begin{tabular}{
    p{0.55\columnwidth}
    >{\centering\arraybackslash}p{0.35\columnwidth}
}
\toprule
\toprule
Parameter & Value \\
\midrule 
\multicolumn{2}{l}{Stellar parameters} \\
$M_{\rm s}$ ($M_\odot$) 
    & $1.467^{+0.079}_{-0.081}$ \\
$R_{\rm s}$ ($R_\odot$) 
    & $1.744^{+0.045}_{-0.042}$ \\
$T_{\rm eff}$ (K) 
    & $6366^{+92}_{-90}$ \\
$\log \singlestoryg$ (dex) 
    & $4.122^{+0.016}_{-0.020}$ \\
$[\mathrm{Fe/H}]$ (dex) 
    & $0.204^{+0.093}_{-0.096}$ \\
\multicolumn{2}{l}{Planetary and orbital parameters} \\
$M_{\rm p}$ ($M_{\rm Jup}$) 
    & $0.910 \pm 0.042$ \\
$P$ (d) 
    & $1.80988158 \pm 3\times10^{-7}$ \\
$a/R_{\rm s}$ 
    & $4.078^{+0.060}_{-0.083}$ \\
$i$ (deg) 
    & $88.5^{+1.0}_{-1.2}$ \\
$K_{\rm RV}$ ($\mathrm{m\,s^{-1}}$) 
    & $117.5 \pm 3.2$ \\
\bottomrule 
\end{tabular}
\begin{tablenotes}[flushleft]
\item \textbf{Notes.}All parameter values are adopted from \citet{Fu2021}.

\end{tablenotes}
\end{table}

\begin{table}[!t]
\centering
\begin{threeparttable}
\caption{Instrumental characteristics and expected noise performance adopted for the WASP-76~b phase curve simulations.}
\label{table:instruments}

\begin{tabularx}{\columnwidth}{lYYl}
\toprule
\toprule
Channel & Wavelength & Cadence & Noise \\ 
        & ($\mu$m)   & (minutes) & (ppm) \\ 
\midrule 
JWST\tnote{a} \\[0.5ex]
NRS1 & 2.87 -- 3.72 & 0.136 & 32.6 \\
NRS2 & 3.82 -- 5.18 & 0.136 & 107  \\ 
\addlinespace [-0.001ex]

\textit{Ariel}\tnote{b} \\[0.5ex]
NIRSpec            & 1.10 -- 1.95 & 0.097 & 555 \\
AIRS Channel 0 (1) & 1.95 -- 2.76 & 1.0   & 291 \\ 
AIRS Channel 0 (2) & 2.76 -- 3.90 & 1.0   & 416 \\ 
AIRS Channel 1     & 3.90 -- 7.80 & 0.444 & 676 \\

\bottomrule
\end{tabularx}
\begin{tablenotes}
\item[a] Noise values are estimated using the JWST ETC (\url{https://jwst.etc.stsci.edu/}).
\item[b] Noise values are estimated using \texttt{ArielRad} \citep{mugnai_arielrad}.
\end{tablenotes}
\end{threeparttable}
\end{table}

For the planetary radius, we adopted the wavelength-dependent values reported by \citet{Fu2021}, selecting the closest available wavelength to each simulated JWST and \textit{Ariel} passband. Because these measurements assume a spherical planetary geometry and do not account for tidal or rotational distortions, we converted them to equivalent volumetric radii assuming the injected fiducial Love number. The resulting volumetric radii adopted for each instrument configuration are listed in Table~\ref{table:priors}. 

To ensure a realistic out-of-eclipse baseline in the forward models, the atmospheric thermal emission parameters ($F_{\rm max}$ and $F_{\rm min}$) and the longitudinal hotspot offset ($\delta$) were anchored to published phase curve observations of WASP-76 b. Specifically, we adopted the values reported by \citet{Demangeon2024} for their empirical `Cos' model, which describes the phase curve modulation using a simple cosine function and was fitted to \textit{Spitzer}/IRAC 4.5~$\mu$m observations. These values were adopted consistently across all simulated JWST and \textit{Ariel} channels to provide an observationally motivated reference phase curve. Although the planetary dayside and nightside fluxes and the hotspot offset are generally expected to vary with wavelength, our simulated observations are restricted to the near- and mid-infrared. Adopting a common empirical baseline therefore avoids introducing additional assumptions about wavelength dependence while providing a representative approximation for the present analysis.

\subsection{Instrument configurations}
\label{sec:instrument_config}

\subsubsection{JWST configuration}
\label{sec:instrument_config_jwst}

We simulated observations with the JWST NIRSpec instrument in the G395H mode, covering the wavelength range 2.87--5.18 $\mu$m. This instrument provides high spectral resolution and excellent photometric precision, making it well suited to detecting the subtle photometric signatures associated with planetary shape deformation. We modelled the expected precision of these data using the JWST Exposure Time Calculator (ETC) and the instrument setup adopted in GO-5268 (PI: J.~Wardenier)\footnote{\url{https://www.stsci.edu/jwst-program-info/download/jwst/pdf/5268/}}. Specifically, we used SUB2048 with 8 groups per integration using the NRSRAPID readout pattern. The G395H mode splits the spectrum onto two detectors, NRS1 and NRS2. Hence, we modelled two light curves for this setup. The properties of the simulated JWST channels are listed in Table~\ref{table:instruments}. 

\subsubsection{\textit{Ariel} configuration}

We simulated observations utilising \textit{Ariel}'s primary infrared science suite, specifically focusing on the Near-Infrared Spectrometer (NIRSpec) and the two \textit{Ariel} InfraRed Spectrometer (AIRS) channels. The expected data quality was modelled using \texttt{ArielRad} \citep{mugnai_arielrad}, the mission's official radiometric tool. The cadence was set such that each channel reached 80\% of the full well-depth. We analysed these data at the Tier 1 resolution, which splits AIRS Channel~0 into two light curves, while white-light curves are used for NIRSpec and AIRS Channel~1 . The properties of the simulated \textit{Ariel} channels are listed in Table~\ref{table:instruments}. 

\subsection{Retrieval technique}

To recover the injected planetary parameters from the synthetic data, we implemented a Bayesian retrieval framework using dynamic nested sampling through the \texttt{dynesty} package \citep{Speagle2020}. The log-likelihood function evaluated the residuals between the simulated observations and the corresponding forward models. This approach followed the methodology of \citet{Akinsanmi2024_theory}, but was adapted to the multi-channel observing architectures of \textit{Ariel} and JWST. In particular, rather than adopting generic noise levels for the simulated observations, we explicitly incorporated realistic instrument-specific photometric uncertainties calculated for the WASP-76\,b system into the likelihood function for each spectroscopic channel.

In our framework, the fluid Love number, $h_{2f}$, was treated as a free parameter and sampled from an uninformative uniform prior, $\mathcal{U}(0.0, 2.5)$. Because the JWST and \textit{Ariel} observing architectures provide simultaneous coverage across multiple independent spectroscopic channels, we performed both individual and joint retrievals for each mission. For the joint retrievals, purely geometric and orbital parameters, namely the scaled semi-major axis ($a/R_{\rm s}$), orbital inclination ($i$), planet-to-star mass ratio ($M_{\rm p}/M_{\rm s}$), and fluid Love number ($h_{2f}$), were shared across all wavelength channels. In contrast, wavelength-dependent atmospheric and stellar parameters, including the planetary fluxes ($F_{\rm d}$, $F_{\rm n}$), hotspot offset ($\delta$), stellar ellipsoidal variation amplitude ($A_{\rm EV}$), limb-darkening coefficients ($c$, $\alpha$), and volumetric planetary radius ($R_{\rm v}$), were treated as independent free parameters in each channel. The priors adopted in the retrieval framework are summarised in Table~\ref{table:priors}. The orbital period was fixed to the value reported by \citet{Fu2021}  (Table~\ref{table:wasp76b_params}) and used only to set the sampling; retrievals were performed in orbital phase.

For the limb-darkening coefficients, we adopted Gaussian priors with standard deviations of 0.01, following the approach of \citet{Akinsanmi2024_theory}. Their study shows that broadening the limb-darkening priors has a stronger impact on optical deformation measurements and comparatively little impact on infrared phase-curve constraints when the deformation signal remains prominent outside transit. Furthermore, \citet{Akinsanmi2024_wasp12} find broadly consistent deformation-detection significances for WASP-12\,b under alternative treatments of limb darkening. For the Doppler-beaming amplitude ($A_{\rm DB}$), we adopted Gaussian priors with standard deviations of 0.005\,ppm. While propagating the uncertainty directly from the observed radial velocity semi-amplitude of WASP-76\,b (Eq.~\ref{eq:A_DB}) yields a wider distribution with $\sigma=0.02$\,ppm, independent test retrievals confirmed that this difference has no measurable impact on the recovered $h_{2f}$ posteriors. Given that the predicted Doppler-beaming amplitude is below 1\,ppm, and its $\sin\phi$ modulation is orthogonal to the symmetric $\cos(2\phi)$ tidal-deformation signal, the narrower prior does not meaningfully impact the retrievals.

For the transit-only retrievals, we restricted the analysed time series to the vicinity of the primary transit in order to minimise the influence of large-scale orbital phase variations. Because the parameters governing these variations, specifically $F_{\rm d}$, $F_{\rm n}$, $\delta$, and $A_{\rm EV}$, cannot be independently constrained from transit observations alone, we adopted informative priors on these quantities. This setup reflects a realistic observational scenario in which such parameters are either constrained by previous phase curve observations or effectively removed through baseline detrending.

\subsection{Retrieval results for WASP-76 b}

\begin{figure*}
    \centering
    \includegraphics[width=\textwidth]{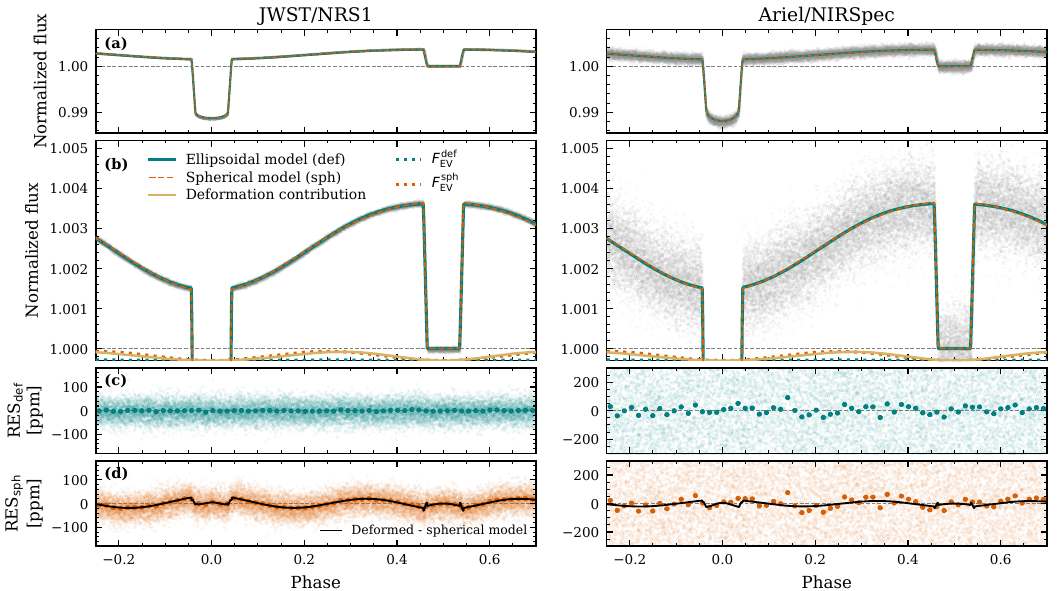}
    \caption{Representative simulated full-orbit phase curves of WASP-76\,b for JWST/NRS1 (left) and \textit{Ariel}/NIRSpec (right), comparing deformed- and spherical-planet fits. Panels a: Simulated data and phase-curve models. Panels b: Zoomed-in view of the phase curves with the fitted models overplotted. The solid cyan and red dashed curves represent the phase-curve models for the deformed and spherical planets, respectively. The dotted curves at the bottom of this panel show the corresponding stellar ellipsoidal variation components ($F_{\rm EV}$), while the yellow curve represents the contribution of tidal deformation to the total phase curve. Panels c and d show the residuals relative to the deformed (cyan) and spherical (red) planet models, respectively. In panel d, the black curve shows the difference between two models, highlighting the component of the phase-curve signal that cannot be fully reproduced by the spherical model. The simulated data show one noise realisation for each instrument.}
    \label{fig:phasecurve_models}
\end{figure*}

Our analysis evaluates the ability of JWST and the forthcoming \textit{Ariel} mission to detect planetary shape deformation and constrain the fluid Love number, $h_{2f}$. A key objective is to quantify the extent to which continuous full-orbit phase curve observations improve the recovery of planetary deformation relative to transit observations alone.

Figure~\ref{fig:phasecurve_models} illustrates the effect of fitting a spherical-planet model to a synthetic phase curve of a tidally deformed planet. The figure shows representative simulated observations of WASP-76\,b with JWST/NRS1 (left) and \textit{Ariel}/NIRSpec (right). The corresponding spherical-planet fits use the same prior distributions as the deformed-planet model (Table~\ref{table:priors}), but with tidal deformation switched off and a broader prior on the stellar ellipsoidal variation amplitude, $A_{\rm EV}\sim\mathcal{U}(0,1000)$~ppm. The spherical model therefore has additional freedom to compensate for the missing deformation signal through the stellar ellipsoidal component. For both the JWST and \textit{Ariel} simulated observations, the residuals of the spherical-planet fits show a clear modulation attributable to tidal deformation that cannot be reproduced by the spherical-planet model. The spherical fits favour substantially larger $A_{\rm EV}$ values in an attempt to reproduce the additional planetary modulation, yielding $A_{\rm EV}\sim116$~ppm and 114~ppm for JWST and \textit{Ariel}, respectively, compared with the injected value of $\sim8$~ppm. However, the stellar ellipsoidal variation cannot fully reproduce the deformation contribution because the latter has a different phase dependence from $F_{\rm EV}$, with peaks occurring at different phases and a shorter inter-peak interval. As a result, systematic structure remains in the residuals of the spherical-planet fits, while the deformed model reproduces the phase-dependent structure of the simulated data. This comparison illustrates how the additional out-of-transit information provided by continuous full-orbit phase curve observations helps distinguish planetary deformation from other phase-curve components and reduces the degeneracies that limit transit-only constraints on $h_{2f}$.

\subsection{Detectability with JWST}

\begin{figure}[!t]
\centering
\begin{subfigure}{\columnwidth}
\centering
\includegraphics[width=0.99\linewidth]{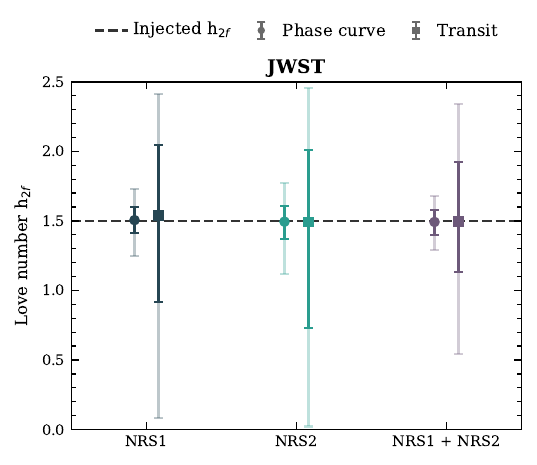}
\end{subfigure}
\begin{subfigure}{\columnwidth}
\centering
\includegraphics[width=0.99\linewidth]{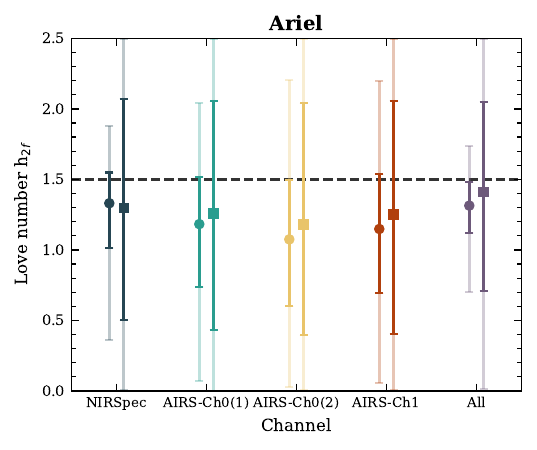}
\end{subfigure}
\caption{Retrieved fluid Love number ($h_{2f}$) from simulated JWST (top) and \textit{Ariel} (bottom) observations of \mbox{WASP-76~b}. Circular markers show constraints from full phase curve observations, while square markers show constraints from transit-shape deformation alone. For JWST, results are shown for NRS1, NRS2, and their joint analysis (NRS1 + NRS2); for \textit{Ariel}, individual spectroscopic channels and a joint fit to all four modelled bands (All) are shown. Marker positions indicate the posterior median, with dark and light error bars denoting the $1\sigma$ and $3\sigma$ credible intervals, respectively. The horizontal dashed line marks the injected value, $h_{2f}=1.5$.}
\label{fig:h2f_retrievals}
\end{figure}

Our simulations indicate that JWST can provide precise constraints on the fluid Love number of WASP-76\,b from a single observing visit, although the adopted observing strategy remains the dominant factor governing the final precision. The retrieved constraints on $h_{2f}$ from simulated JWST/NIRSpec-G395H observations are shown in Fig.~\ref{fig:h2f_retrievals} for the NRS1 and NRS2 detectors individually and for their joint combination.

Although the high photometric precision of JWST allows a weak constraint on $h_{2f}$ from the primary transit alone, the resulting transit-only constraint remains relatively imprecise. As illustrated by the square markers in Fig.~\ref{fig:h2f_retrievals}, the absence of a continuous out-of-transit baseline results in broad posterior distributions. For the individual transit-only retrievals, NRS1 and NRS2 yield poor absolute precisions, with posterior widths of $\sigma_{h_{2f}} \approx 0.56$ and $\sigma_{h_{2f}} \approx 0.64$, respectively. The corresponding $3\sigma$ credible intervals remain broad, indicating that transit morphology alone provides only limited constraints on planetary deformation. Combining NRS1 and NRS2 improves the precision, reducing the posterior width to $\sigma_{h_{2f}} \approx 0.40$, with a recovered median value of $h_{2f} \approx 1.5$.

The large uncertainties obtained in the transit-only scenario arise primarily from geometric parameter degeneracies. Despite the adoption of informative priors on the out-of-transit phase curve parameters, strong correlations persist between $h_{2f}$ and key transit-shape parameters, including the scaled semi-major axis, orbital inclination, and volumetric planetary radius \citep[e.g.][]{Barnes2003, Correia2014a, Akinsanmi2019, Hellard2019}. Because planetary deformation manifests primarily through subtle ingress and egress distortions, the transit signal alone cannot uniquely distinguish between varying degrees of deformation and small changes in orbital geometry. In addition, the deformation signature remains partially degenerate with stellar limb darkening. Together, these effects broaden the posterior distributions and limit the precision with which $h_{2f}$ can be recovered from transit observations alone.

In contrast, full-orbit phase curve observations capture the variation in the projected area of the tidally distorted planet throughout its orbit, enabling a substantially more robust measurement of the Love number (Fig.~\ref{fig:h2f_retrievals}). The phase curve retrievals provide strong constraints, recovering median values of $h_{2f}=1.51$ (NRS1), $1.49$ (NRS2), and $1.49$ (NRS1+NRS2), with precisions of $\sigma_{h_{2f}}\approx0.09$--$0.12$. These correspond to a relative uncertainty of approximately $6\%$ in $h_{2f}$ for the combined retrieval, and $\sim18\%$ in the corresponding potential Love number $k_{2f}$ (assuming $h_{2f}=k_{2f}+1$). In all cases, the injected value is accurately recovered and the resulting posterior distributions remain tightly constrained even at the $3\sigma$ level. 

The precision of the retrieved Love number also depends on the constraints on the system parameters entering the deformation model. In particular, the phase-curve deformation signal constrains the combination $h_{2f} M_{\rm s}/M_{\rm p}$ (Eq.~\ref{eq:Adef}), meaning that uncertainties in the planet-to-star mass ratio propagate directly into the inferred $h_{2f}$. The $\sim6\%$ precision obtained from the joint JWST retrieval is therefore conditional on the adopted Gaussian prior on $M_{\rm p}/M_{\rm s}$ (Table~\ref{table:priors}), which has a fractional uncertainty of approximately $6.8\%$. As phase curve observations become increasingly precise, the uncertainty in the mass ratio can become a limiting factor in the precision achievable for Love-number measurements. This highlights the importance of obtaining precise and independent mass-ratio constraints to fully exploit phase curve observations for measuring planetary Love numbers.

The broad wavelength coverage of NIRSpec-G395H enables a powerful joint retrieval across both detectors. Because the deformation signal is intrinsically achromatic, whereas the atmospheric emission is strongly wavelength dependent, simultaneously fitting both spectroscopic channels helps mitigate degeneracies between planetary shape deformation and atmospheric heat redistribution.

Recent interior-structure studies suggest that measurements of the Love number $k_{2f}$ with precisions better than $\sim 40\%$ are required to provide meaningful constraints on planetary interiors \citep[e.g.][]{vanDijk2025}. The $\sim 18\%$ precision obtained for $k_{2f}$ from a single simulated JWST phase curve of WASP-76\,b therefore exceeds this threshold by a considerable margin. This performance is markedly superior to that achieved with transit-only observations and, as shown below, also surpasses that expected from prospective multi-visit \textit{Ariel} observing campaigns.

\subsection{Detectability with \textit{Ariel}}
\label{sec:results_ariel}

Our retrieval simulations of WASP-76 b for \textit{Ariel} suggest that continuous full-orbit phase curve coverage would be required to obtain robust constraints on planetary deformation from photometric observations. The retrieved constraints on the fluid Love number for different observing strategies are shown in Fig.~\ref{fig:h2f_retrievals}. Similar to the JWST simulations, retrievals based solely on the primary transit struggle to constrain $h_{2f}$ effectively. As illustrated by the square markers in Fig.~\ref{fig:h2f_retrievals}, the transit-only retrievals yield broad posterior distributions across all individual channels, with absolute $1\sigma$ uncertainties of around 0.8. The corresponding $3\sigma$ credible intervals remain extremely broad, offering no meaningful constraint on the planetary shape. For the joint retrieval combining all instruments, the transit-only case yields an uncertainty of $\sigma_{h_{2f}} \approx 0.67$, with a retrieved median value of $h_{2f} \approx 1.41$.

Future \textit{Ariel} phase curve observations should be capable of capturing the changing projected area of the deformed planet, effectively breaking these degeneracies. For a simulated phase curve observation of WASP-76 b with \textit{Ariel} (circle markers in Fig.~\ref{fig:h2f_retrievals}), the individual instruments provide constraints of varying precision. NIRSpec delivers the strongest independent constraint, with an uncertainty of $\sigma_{h_{2f}} \approx 0.27$, while the AIRS channels yield broader constraints of $\sigma_{h_{2f}} \approx 0.39$ for AIRS Channel 0 (1), $\sigma_{h_{2f}} \approx 0.45$ for AIRS Channel 0 (2), and $\sigma_{h_{2f}}\approx 0.42$ for AIRS Channel 1. Combining all channels significantly improves the precision of the retrieval, reducing the uncertainty to $\sigma_{h_{2f}} \approx 0.18$. Evaluated against the injected Love number, this combined retrieval corresponds to a relative precision of approximately 12\% on $h_{2f}$ and 36\% on $k_{2f}$. We nevertheless find a small systematic offset in the retrieved median value. This residual bias is consistent with remaining degeneracies between the planetary deformation signal, the stellar ellipsoidal variation, and the inferred volumetric radius (Appendix~\ref{app:cornerplots}), similar to those identified by \citet{Akinsanmi2024_theory}. As discussed below, these residual biases diminish as the signal-to-noise ratio increases.

\begin{figure}[!t]
\centering 
\includegraphics[width=0.5\textwidth]{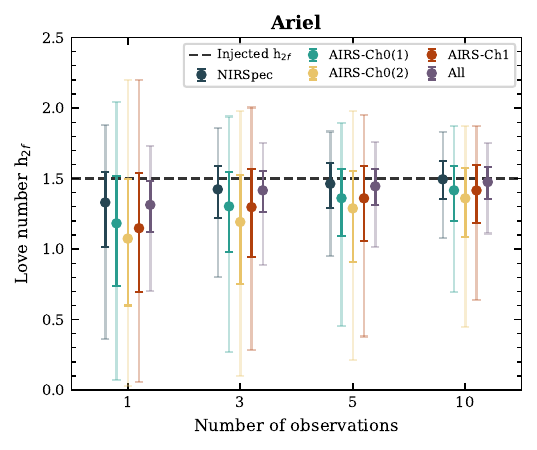}
\caption{Retrieved fluid Love number ($h_{2f}$) as a function of the number of simulated \textit{Ariel} phase curve observations of \mbox{WASP-76~b}. Constraints are evaluated for 1, 3, 5, and 10 stacked visits to illustrate the expected improvement in measurement precision. Results are displayed for individual \textit{Ariel} channels alongside a joint fit incorporating all four of these modelled bands (All). Marker positions indicate the median posterior value, with dark and light error bars denoting the $1\sigma$ and $3\sigma$ credible intervals, respectively. The horizontal dashed line marks the injected value, $h_{2f} = 1.5$.}
\label{fig:ariel_multi_visit}
\end{figure}

To investigate the impact of increased photon collection, we additionally simulated stacked phase curve observations, the results of which are shown in Fig.~\ref{fig:ariel_multi_visit}. Increasing the number of visits from one to ten progressively reduces the posterior width for all instruments and improves the consistency between channels. For example, after three stacked visits, the uncertainty for the combined retrieval drops to $\sigma_{h_{2f}} \approx 0.15$ (corresponding to a relative precision of roughly 29\% on $k_{2f}$). By ten stacked visits, the precision of the independent constraints tightens significantly across all channels, reaching $\sigma_{h_{2f}} \approx 0.13$ for NIRSpec, and ranging from $\sigma_{h_{2f}} \approx 0.19$ to $0.24$ for the individual AIRS channels. Alongside these improvements in precision, the retrieved posterior medians simultaneously converge towards the injected Love number. This steady enhancement in both precision and accuracy demonstrates how additional photon collection efficiently mitigates the residual geometric degeneracies.

Recalling the $\sim 40\%$ precision threshold required to meaningfully constrain planetary interior models \citep[e.g.][]{vanDijk2025}, transit-only observations are unable to reach the precision necessary to overcome the inherent geometric degeneracies. Full-orbit phase curve observations therefore appear essential for \textit{Ariel} to measure planetary deformation at a level that can inform interior models. Our simulations indicate that a single joint \textit{Ariel} phase curve observation of WASP-76 b could achieve a precision of approximately 36\% on $k_{2f}$, satisfying this requirement. Although combining multiple phase curves further improves the constraints, the substantial observing time required for a multi-visit campaign may limit its practicality when deformation measurements are the primary science objective. A single continuous full-orbit observation, therefore, emerges as an efficient \textit{Ariel} observing strategy for highly deformed planets such as WASP-76 b, providing interior-relevant constraints while maintaining efficient use of observing resources.

\section{Discussion}
\label{sec:discussion}

Our analysis demonstrates that both JWST and the future \textit{Ariel} mission are expected to provide the photometric precision required to detect planetary deformation and measure fluid Love numbers for favourable systems. The proposed deformation metric, $\mathcal{D}$, provides a straightforward, physically motivated criterion to prioritise these targets. Ultra-hot Jupiters exhibiting the highest $\mathcal{D}$ values, such as WASP-121\,b, WASP-33\,b, WASP-76\,b, WASP-12\,b, WASP-19\,b, HAT-P-32\,b, KELT-9\,b, WASP-103\,b, KELT-20\,b, and HAT-P-70\,b, emerge as prime candidates for dedicated full-orbit phase curve observing programmes.

This metric is explicitly constructed to quantify the intrinsic detectability of the tidal deformation signal, isolating the physical amplitude of the signature from observational overheads. Capturing a complete planetary phase curve requires continuous monitoring over at least one orbital period. However, entangling the physical signal strength with the logistical constraints of specific observatories would obscure the intrinsic scientific potential of a given system. Extracting the fluid Love number, $h_{2f}$, remains one of the few empirical pathways to break deep degeneracies in planetary interior models, such as constraining core mass fractions and bulk metallicities, thereby motivating the substantial telescope time investment required for observations of high-$\mathcal{D}$ targets.

While balancing observing time against scientific return is a practical necessity for flagship facilities like JWST and \textit{Ariel}, systems with the largest values of $\mathcal{D}$ are, in many cases, also highly suited to near-term observational campaigns. As demonstrated by our highest-ranked systems (Table~\ref{table:candidates}), the leading candidates predominantly possess very short orbital periods, with the majority orbiting in less than two days. Furthermore, as illustrated in Fig.~\ref{fig:candidates} and Fig.~\ref{fig:deformation_metric_dependence}, planetary inflation plays a major role in shaping the detectability threshold, allowing highly inflated planets to remain attractive targets across a broad range of host star magnitudes. Consequently, $\mathcal{D}$ tends to favour systems that are also observationally accessible for continuous phase curve monitoring.

Nevertheless, the extreme nature of these prime candidates occasionally introduces specific astrophysical considerations. Because the strongest tidal deformations naturally occur in highly irradiated ultra-hot Jupiters, several of the top-ranked targets orbit early-type, rapidly rotating, or active host stars. For instance, WASP-33 exhibits $\delta$ Scuti pulsations\citep[e.g.][]{Collier_wasp33, vonEssen_wasp33}, while KELT-9 features pronounced gravity darkening driven by rapid rotation \citep[e.g.][]{Gaudi_kelt9, Ahlers_kelt9}. Although such intrinsic stellar variability can introduce complex photometric modulations that overlap with the symmetric $\cos(2\phi)$ planetary deformation harmonic, these stellar effects can often be modelled and mitigated, particularly by leveraging the broad wavelength coverage provided by JWST and \textit{Ariel}. Consequently, the high-$\mathcal{D}$ ranking serves as an optimised starting point; observers can readily cross-match these prime targets with known host star behaviour to ensure reliable signal extraction in future surveys.

Translating this theoretical metric into practical expectations, our end-to-end simulated retrievals for WASP-76 b demonstrate that the adopted observational strategy is critical. Specifically, while simulated isolated transit observations with JWST and \textit{Ariel} can provide loose constraints on the Love number of the planet, they remain limited by geometric degeneracies with the semi-major axis, orbital inclination, and volumetric planetary radius. In contrast, continuous full-orbit phase curves significantly reduce these degeneracies. For JWST, its superior photon-collecting area allows a simulated phase curve observation of WASP-76\,b to yield a tight relative precision of roughly 18\% on the potential Love number $k_{2f}$. This prospective precision comfortably exceeds the $\sim 40\%$ threshold required to meaningfully differentiate between competing interior structure models, such as those with varying core masses or enriched heavy-element envelopes \citep{vanDijk2025}. Simulated \textit{Ariel} observations of WASP-76 b suggest that comparable constraints may also be achievable; our simulations show that a single visit can achieve a prospective relative precision of approximately 36\% in $k_{2f}$. While these specific precision values are inherently tied to the large signal amplitude and highly favourable signal-to-noise ratio of WASP-76\,b, these predictions suggest that stacking multiple \textit{Ariel} phase curves to further reduce the uncertainties may not be strictly necessary to obtain meaningful interior constraints for similarly favourable, highly deformed targets.

Such constraints, obtained from a single phase curve observation, would represent a highly efficient use of observing time while providing direct insight into planetary interiors. Measuring $h_{2f}$ offers an empirical probe of the planetary mass distribution, a fundamental property that cannot be inferred from atmospheric spectroscopy alone. A high value of $h_{2f}$ ($h_{2f} \to 2.5$) would indicate a weakly centrally condensed interior and therefore a relatively homogeneous mass distribution, potentially consistent with processes such as core erosion or efficient interior mixing. Conversely, a low value ($h_{2f} \to 1.0$) would imply a strongly centrally condensed interior, consistent with the presence of a substantial core. Fluid Love number measurements therefore provide a unique observational avenue for constraining the internal structure and evolutionary history of giant exoplanets.

Despite these promising prospects, our simulations also highlight several observational challenges. The fractional amplitude of the shape-induced planetary deformation is inherently small (typically of the order of tens to hundreds of ppm; \citealt{Akinsanmi2024_wasp12, Akinsanmi2024_theory}) and can exhibit strong degeneracies with other components of the phase curve. Multi-band photometry and high-resolution spectroscopy, both available with JWST and expected from \textit{Ariel}, could help mitigate these effects. Because the geometric deformation signal is achromatic, fitting multiple wavelength channels simultaneously can help distinguish it from the highly chromatic thermal emission. Nevertheless, uncorrected instrumental systematics or stellar activity at the tens-of-ppm level could easily mimic or mask the symmetric $\cos(2\phi)$ harmonic, highlighting the need for precise instrument calibration and baseline stability. The importance of accounting for these effects has been demonstrated in previous high-precision exoplanet studies. For example, \citet{Barros2022} modelled instrumental systematics using multidimensional Gaussian processes in their analysis of the tidal deformation of WASP-103\,b. Similarly, \citet{Sarkar2021} presented an end-to-end simulation framework designed to capture time-correlated noise and time-dependent systematics in exoplanet observations, including applications to the \textit{Ariel} mission. In our case study of WASP-76\,b we accounted for instrument-specific photometric noise estimates but did not explicitly model time-correlated noise. Consequently, our precision estimates should be interpreted in the context of the adopted noise assumptions, with the constraints achievable in actual observations also depending on instrumental stability and the effectiveness of correlated-noise characterisation.

Ultimately, these findings complement and extend previous theoretical frameworks for modelling shape deformation in hot Jupiters \citep[e.g.][]{Correia2014a, Akinsanmi2019, Hellard2019, Akinsanmi2024_theory}. We build on this foundation by introducing the deformation metric to systematically prioritise the most favourable targets and by performing end-to-end simulations that explicitly incorporate realistic instrumental noise estimates. By connecting theoretical predictions of planetary deformation with realistic observing strategies, we demonstrate that high-precision measurements of $h_{2f}$ are immediately within reach.

Given the exceptional precision achievable with a single JWST phase-curve observation, conducting a dedicated survey would be an efficient use of observing time. Rather than requiring exhaustive multi-visit campaigns, an intensive JWST survey targeting the highest-$\mathcal{D}$ planets could comprehensively measure fluid Love numbers for a representative sample of highly deformed planets within a feasible allocation of telescope time. In the coming decade, \textit{Ariel} is expected to complement these efforts, utilising its dedicated mission architecture to extend this emerging parameter space to a substantially broader catalogue of deformed exoplanets.

\section{Conclusions}
\label{sec:conclusions}

Empirical measurements of fluid Love numbers through high-precision photometric phase curves represent a new frontier in the characterisation of exoplanet interiors. By quantifying the geometric amplitude of planetary deformation, these measurements provide direct access to the internal mass distribution, a fundamental property that cannot be inferred from mass and radius measurements alone.

In this work, we introduced the deformation metric, a physically motivated ranking criterion that combines the theoretical amplitude of tidal deformation with the expected instrumental signal-to-noise ratio. Applying this metric to the candidate sample for the \textit{Ariel} mission, we identified a top tier of ultra-hot Jupiters, led by highly irradiated targets such as WASP-121\,b, WASP-33\,b and WASP-76\,b, as the most favourable systems for dedicated deformation studies.

Using WASP-76\,b as a benchmark, we performed end-to-end forward-modelling and retrieval simulations to assess the recoverability of an injected Jupiter-like Love number. Our analysis demonstrates that the observational strategy plays a critical role. While isolated transit observations can provide broad constraints on $h_{2f}$, they do not achieve high precision due to persistent degeneracies with fundamental transit-shape parameters. In contrast, continuous full-orbit phase curves effectively reduce these degeneracies. Our joint retrieval framework shows that a single full-orbit phase curve obtained with JWST/NIRSpec can constrain the fluid Love number $h_{2f}$ with a relative precision of approximately 6\%. Similarly, a single joint phase curve observation with \textit{Ariel} yields robust constraints, achieving a relative precision of approximately 12\% on $h_{2f}$.

These results highlight efficient observing strategies for the near future of exoplanet science. Because a single JWST or \textit{Ariel} phase curve can provide precise constraints on $h_{2f}$, assembling a statistical sample of planetary deformation measurements does not require prohibitive amounts of telescope time. JWST is well positioned to conduct the first dedicated survey of fluid Love numbers by targeting the highest-$\mathcal{D}$ systems, providing the data needed to place systematic constraints on planetary interiors. \textit{Ariel}’s dedicated survey architecture will subsequently enable a wider and more extensive follow-up survey in the coming decade.

Ultimately, studies of tidal deformation will complement traditional interior-structure modelling and atmospheric retrieval analyses by providing additional constraints on planetary composition and evolution. We therefore advocate that future observing strategies prioritise programmes consisting of single, continuous full-orbit phase curve observations of the most favourable ultra-hot Jupiters. Establishing $h_{2f}$ as a primary science objective for these systems has the potential to reduce longstanding degeneracies in interior models and to open a new observational window into the deep interiors of extreme exoplanets.

\begin{acknowledgements}
The authors thank the anonymous referee for their constructive comments and suggestions that helped to improve the manuscript. This project has received funding from the European Research Council (ERC) under the European Union’s Horizon 2020 research and innovation programme (grant agreement no. 101088557, N-GINE). BA acknowledges support by the Australian Government through the Australian Research Council's Discovery Projects funding scheme (project DP250103019). The views expressed herein are those of the authors and are not necessarily those of the Australian Government or Australian Research Council.
\end{acknowledgements}

\bibliographystyle{aa}
\bibliography{ref}

\begin{appendix}

\section{Dependence of the deformation metric on planetary, orbital, and stellar properties}
\label{app:deformation_metric_dependence}

\begin{figure*}
    \centering
    \includegraphics[width=\textwidth]{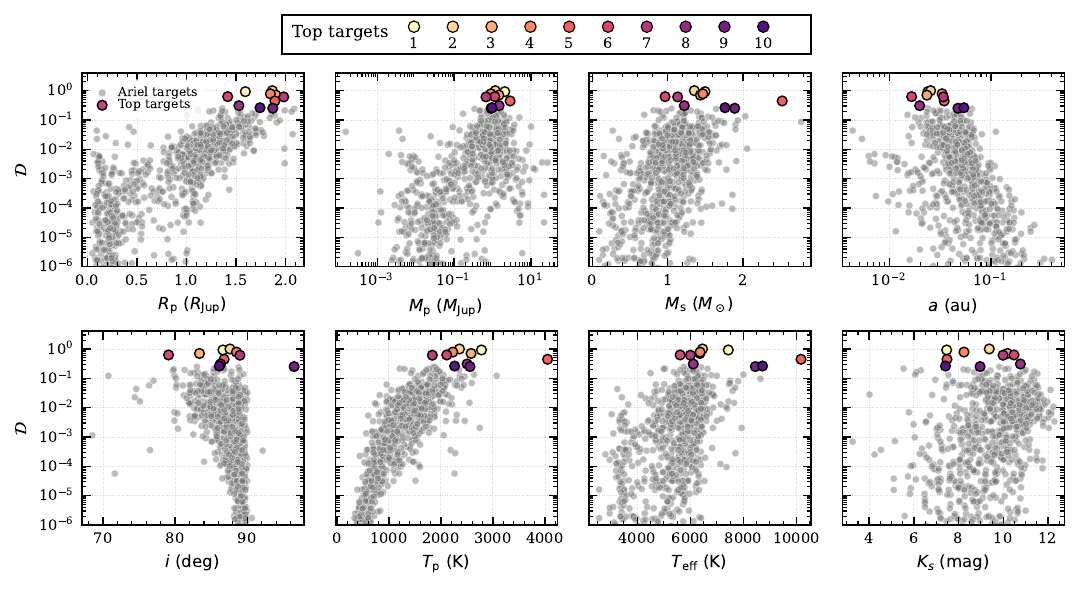}
    \caption{Dependence of the deformation metric, $\mathcal{D}$, on the planetary, orbital, and stellar properties of the \textit{Ariel} target population. From left to right and top to bottom, the panels show $\mathcal{D}$ as a function of planetary radius ($R_{\rm p}$), planetary mass ($M_{\rm p}$), stellar mass ($M_{\rm s}$), orbital separation ($a$), orbital inclination ($i$), planetary equilibrium temperature ($T_{\rm p}$), stellar effective temperature ($T_{\rm eff}$), and host-star $K_{\rm s}$ magnitude. Grey points represent the full \textit{Ariel} target population. The ten highest-ranked targets from Table~\ref{table:candidates} are highlighted and colour-coded according to their ranking, with rank 1 corresponding to the highest value of $\mathcal{D}$.}
    \label{fig:deformation_metric_dependence}
\end{figure*}

Figure~\ref{fig:deformation_metric_dependence} illustrates the dependence of the deformation metric, $\mathcal{D}$, on the main planetary, orbital, and stellar parameters entering its calculation. The grey points show the full \textit{Ariel} target population, while the ten highest-ranked targets identified in Table~\ref{table:candidates} are highlighted and colour-coded according to their ranking. The metric shows strong trends with planetary radius, orbital separation, and planetary equilibrium temperature, reflecting the dependence of the tidal deformation and the expected phase-curve signal on these properties. Additional dependences arise from the planet and stellar masses, orbital inclination, stellar effective temperature, and host-star $K_{\rm s}$ magnitude. Together, these trends illustrate how $\mathcal{D}$ combines the intrinsic tidal deformation of a planet with the observational conditions that determine the detectability of the corresponding phase-curve signal.

\section{Prior distributions and injected parameter values}
\label{app:retrieval_priors}

Table~\ref{table:priors} summarises the injected parameter values used to generate the simulated JWST and \textit{Ariel} phase curve observations of WASP-76\,b, together with the prior distributions adopted in the corresponding retrievals.

\begin{table*}
\centering
\begin{threeparttable}
\caption{Prior distributions and injected parameter values adopted for the simulated phase curve retrievals of \mbox{WASP-76~b}.}
\label{table:priors}
\begin{tabularx}{0.9\textwidth}{llYYYYY}
\toprule
\toprule
Parameter & Distribution & Lower bound & Upper bound & $\mu$ & $\sigma$ & Injected value \\ 
\midrule

\multicolumn{7}{l}{Shared parameters} \\
\addlinespace
$a/R_{\rm s}$               & Uniform & 3.5 & 4.5 & ...      & ...      & 4.073774 \\
$i$ (deg)             & Uniform & 80  & 90  & ...      & ...      & 88.5 \\
$M_{\rm p}/M_{\rm s}$ & Normal  & 0   & 1   & 0.00059 & 0.00004 & 0.00059 \\
$h_{2f}$              & Uniform & 0.0 & 2.5 & ...      & ...      & 1.5 \\

\addlinespace
\multicolumn{7}{l}{JWST instruments} \\
\addlinespace
\multicolumn{7}{l}{NIRSpec - NRS1} \\ 
$R_{\rm v}/R_{\rm s}$            & Uniform & 0.09 & 0.13 & ...     & ...    & 0.11346   \\
$c_{\rm LD}$         & Normal  & 0    & 1    & 0.2843 & 0.01  & 0.2843 \\
$\alpha_{\rm LD}$    & Normal  & 0    & 1    & 0.4350 & 0.01  & 0.4350 \\
$F_{\rm d}$ (ppm)    & Uniform & 2500 & 4500 & ...     & ...    & 3633 \\
$F_{\rm n}$ (ppm)    & Uniform & 1200 & 1700 & ...     & ...    & 1459 \\
$A_{\rm EV}$ (ppm)   & Uniform & 0    & 100  & ...     & ...    & 7.64 \\
$A_{\rm DB}$ (ppm)   & Normal  & ...   & ...   & 0.543  & 0.005 & 0.543 \\
$\delta$ (deg)       & Uniform & -90  & 90   & ...     & ...    & 0.5 \\
\addlinespace

\multicolumn{7}{l}{NIRSpec - NRS2} \\
$R_{\rm v}/R_{\rm s}$          & Uniform & 0.09 & 0.13 & ...     & ...    & 0.11712   \\
$c_{\rm LD}$         & Normal  & 0    & 1    & 0.2201 & 0.01  & 0.2201 \\
$\alpha_{\rm LD}$  & Normal  & 0    & 1    & 0.4571 & 0.01  & 0.4571 \\
$F_{\rm d}$ (ppm)   & Uniform & 2500 & 4500 & ...     & ...    & 3633 \\
$F_{\rm n}$ (ppm)   & Uniform & 1200 & 1700 & ...     & ...    & 1459 \\
$A_{\rm EV}$ (ppm)  & Uniform & 0    & 100  & ...     & ...    & 7.55 \\
$A_{\rm DB}$ (ppm)  & Normal  & ...   & ...   & 0.504  & 0.005 & 0.504 \\
$\delta$ (deg)      & Uniform & -90  & 90   & ...     & ...    & 0.5 \\

\addlinespace
\multicolumn{7}{l}{\textit{Ariel} instruments} \\
\addlinespace
\multicolumn{7}{l}{NIRSpec} \\
$R_{\rm v}/R_{\rm s}$          & Uniform & 0.09 & 0.13 & ...     & ...    & 0.11390   \\
$c_{\rm LD}$         & Normal  & 0    & 1    & 0.5002 & 0.01  & 0.5002 \\
$\alpha_{\rm LD}$  & Normal  & 0    & 1    & 0.4392 & 0.01  & 0.4392 \\
$F_{\rm d}$ (ppm)   & Uniform & 2500 & 4500 & ...     & ...    & 3633 \\
$F_{\rm n}$ (ppm)   & Uniform & 1200 & 1700 & ...     & ...    & 1459 \\
$A_{\rm EV}$ (ppm)  & Uniform & 0    & 100  & ...     & ...    & 8.03 \\
$A_{\rm DB}$ (ppm)  & Normal  & ...   & ...   & 0.787  & 0.005 & 0.787 \\
$\delta$ (deg)      & Uniform & -90  & 90   & ...     & ...    & 0.5 \\
\addlinespace
\multicolumn{7}{l}{AIRS Channel 0 (1) and (2)} \\
$R_{\rm v}/R_{\rm s}$          & Uniform & 0.09 & 0.13 & ...     & ...    & 0.11348 \\
$c_{\rm LD}$         & Normal  & 0    & 1    & 0.3524 & 0.01  & 0.3524 \\
$\alpha_{\rm LD}$  & Normal  & 0    & 1    & 0.4040 & 0.01  & 0.4040 \\
$F_{\rm d}$ (ppm)   & Uniform & 2500 & 4500 & ...     & ...    & 3633 \\
$F_{\rm n}$ (ppm)   & Uniform & 1200 & 1700 & ...     & ...    & 1459 \\
$A_{\rm EV}$ (ppm)  & Uniform & 0    & 100  & ...     & ...    & 7.71 \\
$A_{\rm DB}$ (ppm)  & Normal  & ...   & ...   & 0.597  & 0.005 & 0.597 \\
$\delta$ (deg)      & Uniform & -90  & 90   & ...     & ...    & 0.5 \\
\addlinespace
\multicolumn{7}{l}{AIRS Channel 1} \\
$R_{\rm v}/R_{\rm s}$          & Uniform & 0.09 & 0.13 & ...     & ...    & 0.11713 \\
$c_{\rm LD}$         & Normal  & 0    & 1    & 0.2149 & 0.01  & 0.2149 \\
$\alpha_{\rm LD}$  & Normal  & 0    & 1    & 0.4610 & 0.01  & 0.4610 \\
$F_{\rm d}$ (ppm)   & Uniform & 2500 & 4500 & ...     & ...    & 3633 \\
$F_{\rm n}$ (ppm)   & Uniform & 1200 & 1700 & ...     & ...    & 1459 \\
$A_{\rm EV}$ (ppm)  & Uniform & 0    & 100  & ...     & ...    & 7.54 \\
$A_{\rm DB}$ (ppm)  & Normal  & ...   & ...   & 0.490  & 0.005 & 0.490 \\
$\delta$ (deg)      & Uniform & -90  & 90   & ...     & ...    & 0.5 \\
\bottomrule
\end{tabularx}
\begin{tablenotes}[flushleft]
\item \textbf{Notes.} For normally distributed parameters, $\mu$ and $\sigma$ denote the mean and standard deviation, respectively. System-wide physical and orbital properties (shared parameters) are fitted jointly across all observations. In contrast, wavelength-dependent photometric variables are treated as channel-dependent and fitted independently for each JWST and \textit{Ariel} spectroscopic band. The two \textit{Ariel}/AIRS Channel 0 subchannels share the same prior distributions and injected parameter values.
\end{tablenotes}
\end{threeparttable}
\end{table*}

\section{Representative posterior distributions}
\label{app:cornerplots}

\begin{figure*}[htbp]
\centering 
\includegraphics[width=\textwidth]{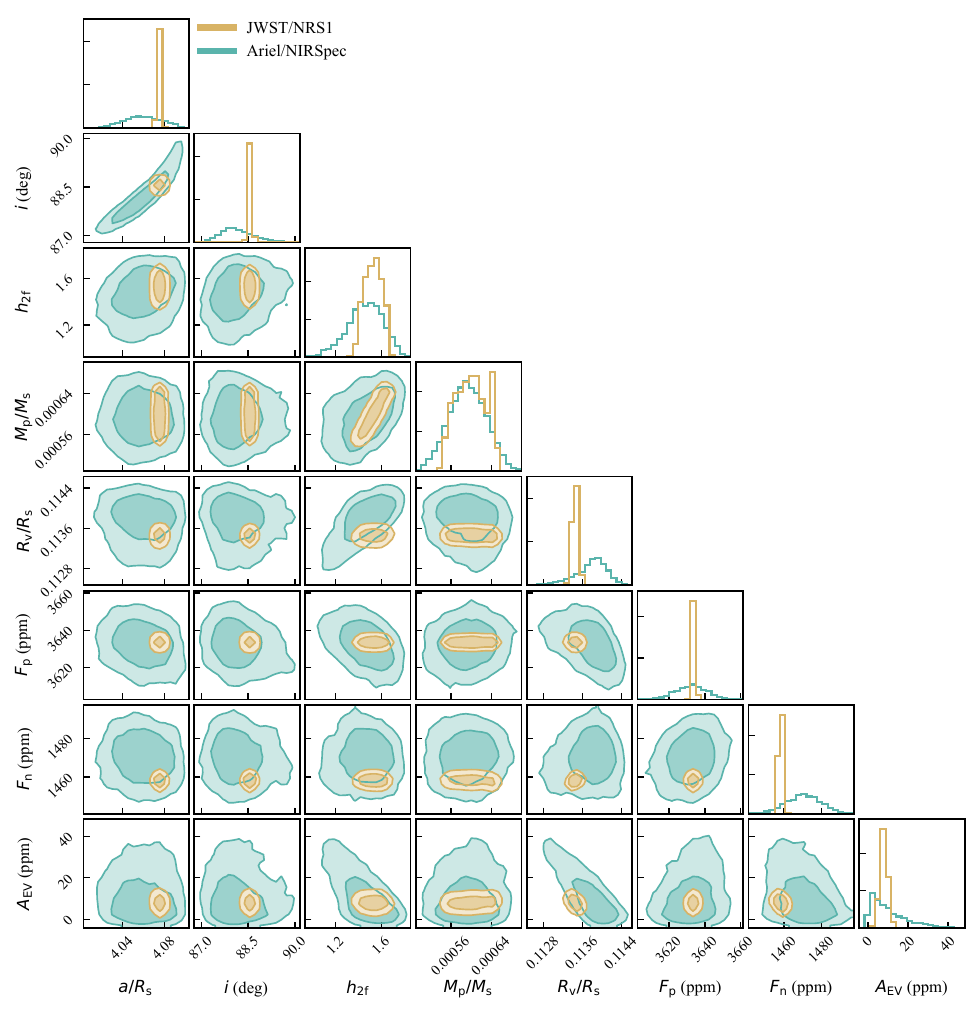}
\caption{Representative posterior distributions for single phase-curve retrievals using JWST/NRS1 (yellow) and \textit{Ariel}/NIRSpec (cyan). The posterior distributions reveal the residual correlations between the fluid Love number $h_{2f}$ and other retrieved system parameters, including the stellar ellipsoidal variation amplitude $A_{\rm EV}$ and the scaled volumetric radius $R_{\rm v}/R_{\rm s}$. These correlations contribute to the small systematic offset in the retrieved median value discussed in Sect.~\ref{sec:results_ariel}. The degeneracies weaken as the signal-to-noise ratio increases (Fig.~\ref{fig:ariel_multi_visit}).}
\label{fig:corner_comparison}
\end{figure*}

Figure~\ref{fig:corner_comparison} presents representative posterior distributions from a typical noise realisation for the simulated single phase curve observations with JWST/NRS1 and \textit{Ariel}/NIRSpec. The corner plots illustrate the parameter correlations that may contribute to the small systematic offset in the retrieved Love number for the simulated \textit{Ariel} observations, as discussed in Sect.~\ref{sec:results_ariel}. Specifically, the planetary fluid Love number exhibits residual covariance with the stellar ellipsoidal variation amplitude ($A_{\rm EV}$) and the inferred scaled volumetric radius ($R_{\rm v}/R_{\rm s}$), reflecting the partial degeneracy between these parameters in the observed phase curve. In contrast, the simulated JWST observations benefit from a higher signal-to-noise ratio, which suppresses these parameter trade-offs and results in a posterior median consistent with the injected value for a single visit. Similar geometric degeneracies have previously been discussed by \citet{Akinsanmi2024_theory}.

Although these correlations produce a modest shift in the posterior median for a single \textit{Ariel} observation, the parameter space remains well constrained. Crucially, the injected value lies well within the $2\sigma$ credible interval, and the resulting precision is sufficient to meet the threshold required for planetary interior-structure studies \citep{vanDijk2025}.

Future implementations could further reduce this degeneracy by computing the stellar ellipsoidal variation amplitude directly from the system geometry (Eq.~\ref{eq:A_EV}), rather than treating $A_{\rm EV}$ as an independent free parameter. In this formulation, $A_{\rm EV}$ would be evaluated self-consistently from the fitted system parameters, including the planet-to-star mass ratio, the scaled semi-major axis, and the orbital inclination, leaving only the passband-dependent ellipsoidal coefficient to be determined. Since $\alpha_{\rm EV}$ depends on the linear limb-darkening coefficient $u$ and the gravity-darkening coefficient $g$, and because $u$ is already included in the phase-curve retrieval, $g$ would replace $A_{\rm EV}$ as the fitted parameter. This reparametrisation enforces a physically motivated relationship between the stellar ellipsoidal variation and the system geometry, and is therefore expected to reduce the covariance between the stellar ellipsoidal variation and the planetary deformation signal.

\end{appendix}

\end{document}